\documentclass[prc,twocolumn,showpacs,preprintnumbers]{revtex4}
\usepackage{bm}
\usepackage{amssymb}
\usepackage{amsmath}
\usepackage{graphicx}

\begin{document}
%%%%%%%%%%%
%%%%%%%%%%%
\title{The black hole spin influence on accretion disk neutrino detection}
%%%
\author{O. L. Caballero}\email{lcaballe@umail.iu.edu}
\author{T. Zielinski}
\affiliation{Department of Physics,
             University of Guelph, Guelph, ON N1G 2W1, Canada }
\author{G. C. McLaughlin} 
\affiliation{Department of Physics,
             North Carolina State University, Raleigh, NC 27695, USA}
\author{R. Surman}
\affiliation{Department of Physics, University of Notre Dame, Notre Dame, IN 46556, USA}

%%%%%%%%%%%
\date{\today}
\begin{abstract}
Neutrinos are copiously emitted from black
hole accretion disks
playing a fundamental role in their evolution, as well as in the production
of gamma ray bursts and r-process nucleosynthesis. The black hole generates a strong gravitational field able to change
the properties of the emerging neutrinos.
We study the influence of the black hole spin on the structure of the neutrino surfaces,
neutrino luminosities, average neutrino energies, and event counts at SuperK. We consider several disk models and provide estimates
that cover different black hole efficiency scenarios. We discuss the influence of the detector's inclination with respect
to the axis of the torus on neutrino properties. We find that tori around spinning black holes have larger luminosities, energies and rates compared to tori around
static black holes, and that the inclination of the observer causes a reduction in the luminosities and detection rates but an increase in the average energies.
\end{abstract}
\smallskip
\maketitle

\section{Introduction}

The physics of stellar phenomena such as, gravitational radiation, 
 the synthesis of heavy elements and short gamma ray 
bursts is closely tied to the emission of neutrinos from black hole (BH) accretion disks 
(AD). These systems can be one possible outcome from BH-neutron star (NS) or NS-NS mergers 
\cite{MacFadyen1999, Taniguchi2005, Lee1999, Rosswog}. In relation to the synthesis of 
elements, given the neutron richness of their progenitors, BH-AD are proposed as good 
candidates for r-process nucleosynthesis \cite{Lattimer1974, Lattimer1976,Surman:2008qf,Fernandez:2014cna}. 
On the other hand, because of the vast amount of neutrinos emitted, neutrino annhilitation 
above the BH could provide the conditions needed to trigger gamma ray bursts 
\cite{Popham1999,SetiawanBHmerger, RuffertGRB-BH,Nakamura:2013bza}. Equally interesting is 
the possibility of neutrino detection from ADs in future and current facilities as has been 
discussed in several works \cite{Nagatakicounts,McLaughlin:2006yy,Caballero:2009ww}.

Fully 3D general relativistic simulations are needed to study the implications of the 
space-time metric and of neutrino cooling on the evolution of BH-NS or NS-NS mergers 
\cite{Foucart:2015vpa,Foucart:2014nda,Deaton:2013sla,Sekiguchi:2015dma,Palenzuela:2015dqa}. 
Such simulations have shown that the evolution of the binary mergers depends on the 
initial binary parameters \cite{Foucart:2014nda}. In particular, in BH-NS mergers the BH 
spin and its alignment characterize the accretion onto the BH 
\cite{Etienne:2008re,Foucart:2010eq}. These simulations offer an important benchmark for 
predictions of neutrino energies and luminosities at infinity. They are, however, 
computationally demanding making simpler models and post-processing good alternatives for 
further studies of neutrino properties.

Good efforts have been devoted to studying BH-AD in the Kerr metric (suitable for spinning BHs) 
and in the Schwarzschild metric (adequate for non-spinning BHs). Some works have studied the 
structure of neutrino cooled disks accreting into rotating BHs, 
\cite{Popham1999,Beloborodovcross}, and find higher neutrino fluxes than those obtained from 
their non-rotating counterparts. Other studies focus on the neutrino annihilation rates and 
find that the deposition energy from neutrino-antineutrino annihilation is larger for 
spinning BHs \cite{Birk07:annhihilation,Zalamea:2008dq,Liu:2015lfa}. Harikae et al 
\cite{Harikae:2009yt} also found considerable changes in the deposition rates when general 
relativistic effects were considered for neutrinos emitted in a long-term collapsar 
simulation. However, the dependence of observable properties, such as neutrino energies and 
event counts, among others, on BH spin has remained unexplored.

Neutrinos in the free streaming regime, i.e. once they have decoupled from the disk matter, 
can reach an observation (or absorption) point. The neutrino properties, e.g. fluxes, 
energies, and luminosities, observed at the point of emission will be different from the 
ones registered by an observer positioned anywhere else. The observed quantities change due 
to neutrino oscillations \cite{Malkus:2012ts,Malkus:2014iqa}, neutrino interactions, and the gravitational field of the BH. For 
the latter, the consideration of the 3D geometry of the source is of particular importance, 
as the relativistic effects (bending of neutrino trajectories and energy shifts) depend on 
the space-time curvature. In previous works \citep{Caballero:2011dw,Surman:2013sya}, we have studied the 
influence of gravity on the emission of neutrinos and the production of heavy elements in 
outflows emerging from BH-AD via a post-processing of merger simulations and following 
neutrino trajectories in strong gravity.

Here we extend our study of the neutrino properties by considering the 
influence of the BH spin and the observer's location. We gain further understanding on the 
structure of the neutrino surface and its correlation to the BH rotation. We provide 
estimates of electron antineutrino detection rates, counts and signal duration at SuperK 
\cite{SK} for a variety of AD models. We also discuss angular dependencies of 
neutrino properties: as the polar inclination of the observer changes, the bending of the 
neutrino trajectories and energy shifts will be affected. If the gravitational field is 
strong the curved trajectories allow detection of neutrinos emitted from regions of the 
neutrino spheres that would be inaccesible otherwise. In section \ref{diskmodel} we discuss 
the models considered, in section \ref{surfaces} we study the influence of the BH spin on 
the structure of the neutrino surfaces, in section \ref{distantobservation} we show our 
results for an observer at infinity, and in section \ref{angulardependence} we consider a 
distant observer at different inclinations. Finally we motive further work and conclude in 
section \ref{conclusions}.

%=========================================================================================
\section{Disk Models}
\label{diskmodel}

Due to the expensive computational cost of fully relativistic 3D simulations of collapsars 
and binary mergers it is worth trying out other approaches to gain physical insight. In this 
paper we use two different disk models. The first one describes a fully general 
relativistic steady state disk as in \cite{Beloborodovcross}. The second one is a two 
dimensional time-dependent hydrodynamical simulation of a torus with a pseudo-relativistic 
potential \citep{Just:2014}.
 
In the first case we extend to 3D the one dimensional model from Chen and Belobodorov 
\cite{Beloborodovcross}, by estimating the vertical structure using a simple hydrostatic 
model that assumes that the gas forming the disk is at equilibrium under the gas and 
radiation pressure. Furthermore we assume axisymmetry. The model is hydrodynamical and uses 
the Kerr metric to account for two values of the BH spin $a=Jc/GM^2$ ($J$ is the total 
angular momentum and $M$ the BH mass). We label these models according to the BH spin: 
``C0'' for $a=0$ and ``Ca'' for $a=0.95$. The mass of the BH is 3$M_\odot$ and the accretion 
rate $\dot{M}$ is constant.

In the second model, from Just et al \citep{Just:2014}, the simulation is set up to be the 
equilibrium configuration of a constant angular momentum axisymmetric torus that emulates 
the final stages of compact binary mergers. The BH mass is 3$M_\odot$ and the BH spins are 
$a=0$ and $a=0.8$. We refer to these tori as J0 and Ja, respectively. The simulation is 
Newtonian and the gravitational effects of the BH are introduced via a modified Newtonian 
potential which is an extension of the Paczynski-Wiita potential \cite{BHpote} to rotating 
BHs \cite{Artemova}. This reproduces the radius of the innermost stable circular orbit in 
the Kerr metric. The structure of the torus evolves in time. To study the effects of the BH 
spin on the neutrino surfaces we use a $t=20$ ms snapshot, when presumably the neutrino 
fluxes are the largest \cite{Surman:2013sya,Caballerogravity}.

%=========================================================================================
\section{Neutrino Surfaces}
\label{surfaces}

The neutrino surfaces, analogous to the neutrino spheres of a protoneutron star (PNS) from a core collapse supernovae, are 
defined by the points above the equatorial plane where neutrinos decouple from the accretion 
torus. This happens when the neutrino 
%optical 
depth $\tau =2/3$. $\tau$ depends on the 
neutrino opacity which is governed by the different scattering processes that neutrinos 
undergo as they diffuse through the torus. As matter is accreted into the BH it becomes 
hotter and nuclei dissociate. Therefore, we consider neutrino scattering from protons, 
neutrons and electrons. For scattering from neutrons and protons we have the charged current 
processes
\begin{equation}
\nu_e +n\rightarrow p+e^-,
\label{nuabsorption}
\end{equation}
\begin{equation}
\label{anuabsorption}
\bar{\nu}_e +p \rightarrow e^+ + n,
\end{equation}
and the neutral current processes $\nu+n \rightarrow \nu+n$ and $\nu+p \rightarrow\nu+p$. We 
also consider elastic scattering from electrons and neutrino-antineutrino annihilation. The 
charged current reactions affect only electron (anti)neutrinos while the other processes 
affect all neutrino flavors. As tau and muon neutrinos scatter through the same processes we 
label these flavors as $\nu_x$. We find proton and neutron number 
densities assuming charge neutrality $Y_e=Y_p$, and the electron number density assuming 
equilibrium of thermal electrons and positrons with radiation. Details on the cross sections 
of the above reactions and on the optical depth calculation can be found in ref. 
\cite{Caballero:2009ww}. We obtain thermally averaged cross sections by assuming a 
Fermi-Dirac distribution for the neutrinos with a temperature equal to the local torus 
temperature and zero chemical potential,
\begin{equation}
\label{averagedcross}
\langle  \sigma_k(E_\nu)\rangle=\frac{\int ^\infty_0\sigma_k(E_\nu)\phi (E_\nu)dE_\nu} {\int^\infty_0\phi(E_\nu)dE_\nu},
\end{equation}
with $\phi(E_\nu)$ the neutrino flux written in terms of the Fermi-Dirac distribution $f_{FD}$,
\begin{equation}
\phi(E_\nu)=\frac{c}{2\pi^2(\hbar c)^3}(E_\nu)^2 f_{FD}.
\label{FDflux}
\end{equation}
This procedure removes the energy dependence of the neutrino surface, as our results have 
been properly energy weighted. Note, however, that the scattering neutrino surfaces differ 
from the effective neutrino surfaces (see e.g. \cite{Perego:2014fma}), with the difference 
being larger for tau/muon neutrinos.

\begin{figure}[ht]
\begin{center}
\includegraphics[trim= 0in 0.1in 0in 0.1in, clip, width=3.1in] {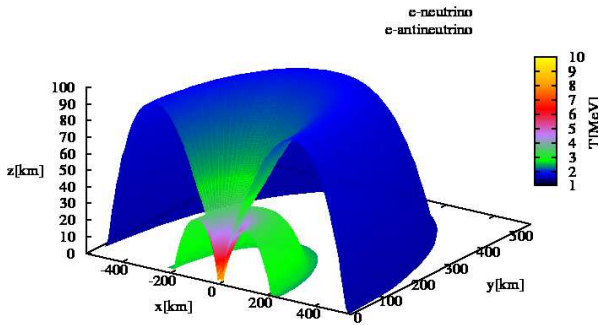}
\caption{(Color on line) Electron neutrino surface (outter) and electron antineutrino surface 
(inner) based on the Ca torus model with a BH spin $a=0.95$, $\dot{M}=5M_\odot$/s and 3$M_\odot$ BH \cite{Beloborodovcross}. 
The color scale shows the corresponding neutrino temperatures.}
\label{nuT}
\end{center}
\end{figure}

Figure \ref{nuT} shows a quarter of the electron neutrino and electron antineutrino surfaces 
corresponding to the Ca model ($a=0.95$) and $\dot{M}=5M_\odot$/s. The color scale 
represents the neutrino temperatures. The hotter neutrinos (yellow) are emitted closer to 
the BH. These will have the biggest contribution to the net neutrino fluxes. Tau/muon neutrino 
surfaces, not shown for clarity, are the hottest ($T_{\nu_x}\approx 10.5$ MeV), followed by 
the electron antineutrino surfaces ($T_{\bar\nu_e}\approx 9$ MeV). The coolest are the electron 
neutrino surfaces which are the farthest from the BH ($T_{\nu_e}\approx$ 8 MeV). This is 
similar to the flavor hierarchy found in the PNS neutrino spheres, and can be understood in 
terms of the reactions eqs. \ref{nuT} and 2. Since the torus is abundant in neutrons, 
electron neutrino absorption on them, eq. \ref{nuabsorption}, will have a more substantial 
effect on the mean free path than electron antineutrino absorption on protons, eq. 
\ref{anuabsorption}, causing electron neutrinos to decouple at the lowest temperatures. 
$\nu_x$ lack these charged current processes and decouple closest to the BH where the 
temperatures are higher.

The geometry and temperatures of the neutrino surfaces for these tori are however different 
from those of a PNS. In the torus, near the BH, due to the gravitational potential, the 
accreted matter forms a funnel with lower densities. Closer to the BH there is an increment 
of the temperature, as matter radiates energy when falling to the BH. The emitted neutrinos 
move through this less dense medium for a shorter time, producing higher temperatures 
compared to a PNS. In the PNS, neutrinos diffuse through a denser medium as compared to a BH-AD, and 
decouple at lower temperatures ($T_{\nu_e}\approx 2.6$, $T_{\bar\nu_e}\approx 4$, and 
$T_{\nu_x}\approx 5$ MeV) \cite{McLaughlin:2006yy}.

\begin{figure}[ht]
\begin{center}
\includegraphics[trim= 1.in 4in 1.5in 0.5in, clip, width=2.5 in]{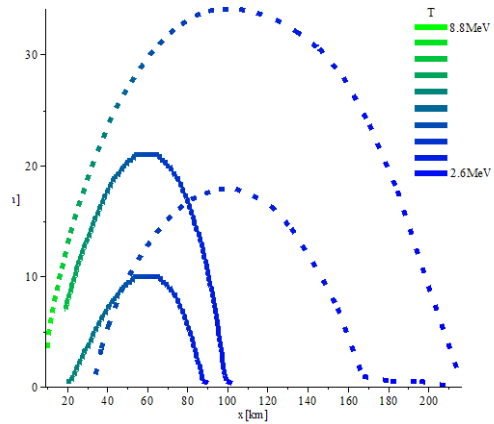}
\caption{Comparison of electron antineutrino surfaces for different BH spins and different disk models.
Dotted lines are for the models C0 ($a=0$) and Ca ($a=0.95$), while solid lines correspond to J0 ($a=0$) and Ja ($a=0.8$). 
The taller lines for each model correspond to the higher spins.}
\label{spin1D}
\end{center}
\end{figure}

Fig. \ref{spin1D} shows a transversal cut of the electron antineutrino surfaces for the two 
models used here: Chen-Beloborodov with $\dot{M}=5M_\odot$/s (dotted lines) and Just et al 
(solid lines). For each model we find the electron antineutrino surfaces for two different 
spin parameters, $a=0$ and $a=0.95$ corresponding to the C0 and Ca models, respectively, and 
$a=0(0.8)$ for J0(Ja) models. In both sets of models, the taller surfaces correspond to the 
higher spin values. The color scale represents the temperature $T_{\bar\nu_{e}}$ at the 
decoupling points. These are higher for spinning BH regardless of the disk model. For the Ca 
model, $T_{\bar\nu_e}$ are as high as 9 MeV, while they are around 4.5 MeV for the C0 case. 
Similarly for the Ja model $T_{\bar\nu_e}$ maximum is 7 MeV while for J0 the highest value 
is around 6 MeV. The BH spin also affects the extension of the disk; the higher the spin the 
larger the neutrino surface of the torus. This, as we explain later, impacts the neutrino 
luminosities and detection rates.
 
Hotter neutrino surfaces for larger spins $a$ are a consequence of the spacetime geometry. 
Matter rotating around a BH will release energy before plunging into it. The radius of the 
marginally stable circular orbit $r_I$, also known as the innermost stable circular orbit or 
ISCO, is the smallest circle along which free particles may stably orbit around a BH. The 
binding energy that can be released increases as $r_I$ decreases, and $r_I$ decreases with 
the increment of the BH spin \cite{Shapiro}.  When the angular momentum per unit mass 
exceeds $r_I$ centrifugal forces will be significant and the matter will circulate around 
the BH. The extra angular momentum will be carried away by viscous stress. The viscous 
heating can then be converted to neutrino luminosity. The extension of the neutrino surfaces 
is also linked to space-time geometry. In Kerr BHs matter is hotter and denser when compared 
to Schwarzschild BHs. Neutrinos scatter in this denser medium taking, therefore, a larger 
distance to decouple which is reflected in larger neutrino surfaces.

%=============================================================================================

\section{Distant Observation}
\label{distantobservation}

Using our results for the neutrino surfaces, and following a similar methodology as in Refs. 
\cite{Caballero:2009ww,Palenzuela:2015dqa}, we can estimate the number of neutrinos emitted 
per sec $f^e$, the emitted luminosities $L^e_\nu$, and average energies $\langle 
E^e_\nu\rangle$ as

\begin{equation}
f^e=\frac{dN}{dt^{e}}=\int dA^{e} dE^{e} \phi^{\mathit{eff}}(E^{e}),
\end{equation}
\begin{equation}
L^{e}_\nu=\frac{dE^{e}}{dt^{e}}=\int dA^{e} dE^{e} E^{e} \phi^{\mathit{eff}}(E^{e}),
\end{equation}
and
\begin{equation}
\langle E^{e}_\nu\rangle=\frac{dE^{e}/dt^{e}}{f^e},
\end{equation}
where the effective flux, i.e. the number of neutrinos emitted per unit energy, per unit area, per second, is
\begin{equation}
\phi^{\mathit{eff}}(E^e)=\frac{1}{4\pi}\int d\Omega^{e}\times\phi(E^{e}),
\label{nufluxes}
\end{equation}
and $\phi(E^e)$ is the Fermi-Dirac flux (eq. \ref{FDflux}).
In the equations above the integral over $dA^{e}$ corresponds to an integral over the 
neutrino surface. Assuming isotropic emission will reduce the integral over $d\Omega^e$ to 
$4\pi$. The transformation of these quantities to an observer located at infinity can be 
done by using the invariance of phase-space density \cite{Caballero:2011dw,Palenzuela:2015dqa},
\begin{equation}
  \frac{1}{c^2}\frac{dN}{d^3x_{o}d^3 p_{o}}=\frac{1}{c^2}\frac{dN}{d^3x_{e}d^3p_{e}},
\end{equation}
with the subindex $o$ denoting observed quantities. The relation between the observed $E^o$ 
and emitted $E^e$ energies is $E^{e}=(1+z) E^{o}$ where $(1+z)$ is the redshift. To a 
distant observer the distances appear stretched roughly by a factor of $(1+z)$, changing the 
areas accordingly. Putting this together we have

\begin{equation}
\frac{dN}{dt^{o}}= \int \frac{1}{(1+z)} \phi(E^e)dE^edA^e,
\end{equation}
\begin{equation}
 \frac{dE^{o}}{dt^{o}}=\int \frac{1}{(1+z)^2} \phi(E^e)E^edE^edA^e,
\end{equation}
and 
\begin{equation}
\langle E^{o}_\nu\rangle=\frac{dE^{o}/dt^{o}}{f^o}.
\end{equation}

The number of neutrinos reaching a detector per second $R$ is given by
\begin{equation}
R=N_T\int_{E_{th}}^{\infty} \phi^{eff}(E^{o})\sigma(E^{o})dE^{o},
\label{rates}
\end{equation}
where $N_T$ is the number of targets in the detector, $E^{o}$ are the neutrino energies
registered, and $\sigma(E^{o})$
is the detector's neutrino cross section.
The effective flux $\phi^\mathit{{eff}} (E^{o})$
 reaching the detector is

\begin{equation}
\phi^{eff}(E^{o})=\frac{1}{4\pi}\int d\Omega^{o}\times\phi(E^{o}_{\nu}),
\label{effflux}
\end{equation}
where $d\Omega^{o}$ is the solid angle that the torus subtends as seen by an observer at the 
detection point, i.e. $d\Omega_{o}=\sin\xi d\xi d\alpha$, where $\xi$, $\alpha$ are 
spherical coordinates in the observer's sky.

We are interested in a distant observer.  For this situation,
 the effective fluxes (eq. \ref{effflux}) will be strongly influenced 
by the energy shifts, while the effects of bending of trajectories, reflected in 
$d\Omega^o$, are smaller. We calculate the energy shifts in the Kerr metric. For the solid 
angle, we follow the null geodesics that travel away from the neutrino surfaces in the 
Schwarzschild metric, in order to ease the calculation without compromising our conclusions. 
Note that if we were interested in an observer that was closer to the BH, a calculation in the Kerr metric would be 
optimal, as null geodesics in the Kerr metric would lead to larger torus areas, seen by the 
observer. For a closer observer estimates of the apparent areas in the Schwarzschild metric 
would be an underestimate as compared with those obtained in the full Kerr metric.

For an observer located at infinity, and in the Schwarzschild metric, it suffices to 
estimate the effect of bending of trajectories by calculating $d\Omega^{o}=b db d\alpha/r^2_{o}$, 
where $b$, the neutrino's impact parameter, is
\begin{equation}
b=\frac{r_+ }{\sqrt{1-r_s/r_+}},
\label{closest}
\end{equation}
$r_s$ is the Schwarzschild radius and $r_+$ is the closest distance of the neutrino 
trajectory to the BH. For a trajectory that goes to infinity $r_+$ equals the emission 
points on the neutrino surface $r_{e}$.

On the other hand, the redshift in the Kerr metric depends on the BH spin $a$, the observer 
and emitter positions, and their relative velocities (see \cite{Caballero:2011dw}). For the 
latter we assume that the emitter and observer are in Keplerian rotation, such that their 
angular velocities are given by
\begin{equation}
\Omega_{o(e)}=\frac{M^{1/2}}{r_{o(e)}^{3/2}+aM^{1/2}},
\label{omegakerr}
\end{equation}
with $r_{o(e)}$ the observer(emitter) distance to the BH. 

We estimate the number of neutrinos per second that could be detected in SuperK assuming 32 
ktons of fiducial volume, with a threshold energy $E_{th}= 5$ MeV \cite{SK}. Here the cross 
section in eq. \ref{rates} corresponds to captures of electron antineutrinos on protons 
\cite{Caballero:2009ww}. We also assume that the torus is at 10 kpc from Earth and that we 
observe it on the $z$-axis, above its equatorial plane. Table \ref{energytable} shows the 
detection rates $R$ observed in SuperK, the observed electron (anti)neutrino luminosities at 
infinity $L^o$, and the corresponding observed average energies for the C and J models. 
Comparing results between the same simulation approach (C0 with Ca, and J0 with Ja) we see 
that these observables are larger for the higher BH spin, a consequence of the conversion of 
the extra rotational energy into thermal energy. It is interesting to note that a PNS would 
produce about 1000 counts/s in SuperK and has a neutrino luminosity of ~$10^{52}$ ergs/s 
\cite{McLaughlin:2006yy}. The tori studied here lead to detection rates greater by a least a 
factor of 10, and luminosities higher by at least one order of magnitude, with the 
difference coming from the hotter neutrino surfaces as discussed in section \ref{surfaces}.

\begin{table}
\begin{center}
\caption{Observed detection rates R (at 10 kpc in SuperK), electron (anti)neutrino luminosities, and energies for the different accretion tori studied here. In all the models the
BH mass is $3M_\odot$}
\begin{tabular}{llllllll}
\hline
  &$a$&R (sec$^{-1}$)&$L^o_{\bar\nu_e}$(ergs/s)&$L^o_{\nu_e}$(ergs/s)& $E^o_{\bar{\nu}_e}$&$E^o_{\nu_e}$ \\
  & &&($\times10^{53}$)&($\times10^{53}$)&(MeV)&(MeV)\\
  \hline
J0&0&15000&2.7 &1.9& 12.7 & 10.3 \\
C0&0&  23900& 4.8& 4.7&10.3 &6.4&\\
Ja&0.8  &  23600& 3.7& 2.4&13.4 & 11 \\
Ca &0.95&  58400& 9.8&6.8 &11.8   & 7.3\\
\hline
\end{tabular} 
\label{energytable}
\end{center}
\end{table}
A full comparison of the microphysics of the two models studied here (Chen-Belobodorov vs 
Just et al.) is not the goal of this study. However, we address the differences between the 
results found in Table \ref{energytable}. Although the luminosities are larger for the C0 
vs the J0 models, the neutrino energies are lower. The same behavior is observed 
between the Ca and Ja tori. The differences can be understood in terms of the dependency of 
the neutrino temperatures with distance to the BH, and the extension of the neutrino 
surfaces. Close to the BH, where most neutrinos are emitted, the temperatures of C0 and J0 
models are similar $\sim$5 MeV (as is also the case for Ca and Ja where T is around tens of 
MeV). However, the surfaces are much larger for the Chen-Belobodorov models when compared to 
the Just et al. ones. This leads to larger luminosities for C0(Ca) despite the higher 
temperatures of J0(Ja). Finally, neutrinos emitted farther from the BH contribute to the 
spectra with lower energies explaining the smaller average energies found for C0(Ca) 
compared to J0(Ja).

A broader picture of the accretion tori can be achieved by placing our post-processed 
results in context with those coming directly from fully relativistic simulations. Recent 
simulations of the long-term evolution of BH-NS mergers with an initial BH spin $a=0.9$ and 
mass ratio of 4 showed the formation of a thick disk of $0.3M_\odot$ and an accretion rate 
of $\sim 2 M_\odot/$s \cite{Deaton:2013sla}. The total initial neutrino luminosity was found 
to be around $10^{54}$ erg/s, dropping to 2$\times 10^{53}$ erg/s 50 ms later. The neutrino 
energies, averaged over time, are 12 and 15 MeV for electron neutrinos and antineutrinos, 
respectively, decreasing at a rate of 1 MeV per 10ms. These are about 1 MeV larger compared 
to our spinning BH results of 11 and 13.4 MeV (Ja model) and larger by about 5 and 3 MeV for 
electron neutrinos and antineutrinos, respectively, for the Ca model.  Note that in ref. 
\cite{Deaton:2013sla} neutrino cooling followed a leakage scheme which neglects neutrino 
absorption.  With an improved neutrino cooling scheme that uses an energy-integrated version 
of the moment formalism F, Foucart et al \cite{Foucart:2015vpa} studied the post-merger 
evolution of a BH-NS binary, focusing on the stage between disk formation and an achievement 
of a quasi-equilibrum phase. Their merger results in a disk of $0.1 M_\odot$ and a BH of $8 
M_\odot$ with spin 0.87. In this case the luminosities are around $5-8\times10^{52}$ erg/s 
for electron neutrinos (the leakage scheme predicted luminosities 30\% larger), $5\times 
10^{53}$ erg/s for electron antineutrinos, while the leakage scheme led to estimated 
energies of 11-13 MeV for electron neutrinos and 14-15 MeV for antineutrinos. 
%It is 
%interesting to note that 
Our results for neutrino luminosities and energies for the spinning 
models in Table \ref{energytable} (in particular for the Ja model), are similar to those 
coming directly from the cited simulations despite the fact that the initial 
%binary 
conditions are different as well as are the neutrino and gravity treatments. This comparison 
is useful to roughly check our results and shows that our ray-tracing technique is a 
reasonable tool 
for predicting neutrino related quantities via post-processing studies.

%=================================================================================================
\subsection{Steady accretion tori counts at SuperK}

To estimate the total number of counts at SuperK we multiply the rates (obtained as Eq. 
\ref{rates}) by the duration of the signal. This time can be calculated as the ratio of the 
energy emitted in the form of neutrinos $E^B_\nu$ to the total neutrino luminosity $\Delta 
t=E^B_\nu/L_\nu$. On the other hand, $E^B_\nu$ depends on the efficiency of converting 
gravitational energy to neutrino energy $\varepsilon=E^B_\nu/E_G$, and so $\Delta 
t=\varepsilon E_G/L_\nu$. The gravitational energy is the rest mass energy $E_G=M_Tc^2$ with 
$M_T$ the torus mass. We estimate the total number of counts for a collection of fully 
relativistic steady state tori, similar to the C models above, with different accretion 
rates. For these models the viscosity parameter is $\alpha=0.1$, the BH mass is 3$M_\odot$, 
and the BH spin takes the values $a=0$ or $a=0.95$. We estimate $\varepsilon$ in two ways. 
For the first, we assume radial accretion, where $\varepsilon=L_\nu/\dot{M}c^2$ 
and then $\Delta t=M_T/\dot{M}$. The second estimate accounts for the fact that the accreted 
matter carries angular momentum and therefore the efficiency depends on the BH spin. 
Then the efficiency $\varepsilon$ is $1-\tilde E$ with 
\begin{equation}
 a=-\frac{4\sqrt{2}(1-\tilde E^2)^{1/2}-2\tilde E}{3\sqrt{3}(1-\tilde E^2)},
\end{equation}
and corresponds to the amount of energy radiated by matter reaching the BH through a series 
of almost circular orbits. $1-\tilde E$ is the maximum of binding energy at the marginally 
stable circular orbit which has a radius $r_{I}$. Thus for a BH of 3 $M_\odot$ with a spin 
$a=0$ we find $r_{I}=26.49$ km and an efficiency $\varepsilon=0.057$. For the spinning BH 
with $a=0.95$, the efficiency is $\varepsilon=0.19$ while $r_{I}=8.91$ km. These 
values are replaced in $\Delta t=\varepsilon M_Tc^2/L_\nu$. Additionally, we take into 
account the time dilation occurring in the vicinity of the BH and our time interval as seen 
at Earth is $\Delta t^{o}=\Delta t(1+z)$,
where this redshift is an estimate taken from 
the overall redshift between the average observed and emitted neutrino energies. Our results 
predict signals lasting $0.14$ to $0.34$ secs for $a=0$, and $0.16$ to $0.39$ sec for 
$a=0.95$ (the estimate for a supernova (SN) is 10 sec).

\begin{figure}
\begin{center}
\includegraphics[trim= 0.in 0.in 0in 0.0in, clip,width=0.45\textwidth]{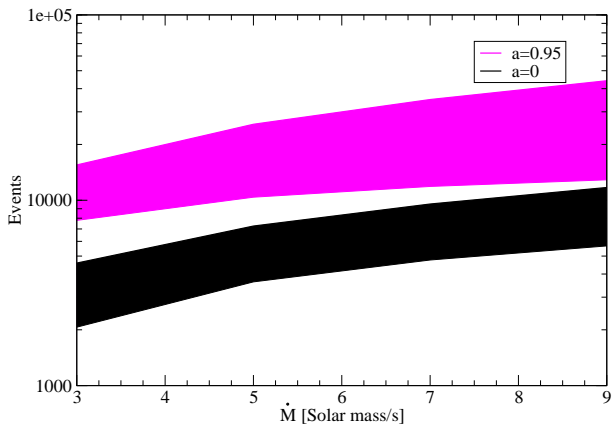}
\caption{Total number of events seen at SuperK as a function of mass accretion rate. The tori models
are 3D extensions of fully relativistic 1D disks. Indicated is the value of the BH spin. The BH mass is $3M_\odot$
in all the cases.}
\label{events}
\end{center}
\end{figure}

Figure \ref{events} shows the total number of counts in SuperK for these tori as a function 
of the accretion rate $\dot{M}$. The magenta band corresponds to $a=0.95$ while the grey one 
to $a=0$. The bands for a fixed value of $a$ are obtained by using our two efficiency 
estimates. The lower limit corresponds to spherical accretion while the upper limit to the 
spin dependent efficiency. The total events from rapidly accreting tori is larger 
because these tori are more massive and so there is more material to be converted in 
neutrino energy. The number of counts can be about an order of magnitude higher for spinning 
BHs than non-rotating BHs in the case of $\dot{M}=3M_\odot$. However this difference is less 
evident as the accretion rate grows. This is because the efficiency calculated in our
first estimate decreases faster with the accretion rate whereas the efficiency in the second 
estimate is constant for all rates. Therefore the corresponding signal duration for $a=0$ in 
the optimistic case will eventually be similar to the pessimistic $a=0.95$ case. If the 
accretion rate is low $\sim 3 M_\odot$/sec then we predict a number of counts close to the 
current SN estimates ($\sim$8000) for a spinning BH and roughly half of the SN events for a 
non-spinning BH. For higher accretion rates $\sim 9 M_\odot$/sec we could have as much as 
four times the number of events of a SN in the most optimistic scenario.

The predictions for the Just et al model are similar to some extent. As this model is time 
dependent, we find the accretion mass rate by calculating the total mass of the torus at 
two different snapshots $t=20$ ms and $t=60$ ms so we get $0.146 M_\odot$/s. The initial 
total mass of the torus is $\sim 0.3 M_\odot$ \cite{Just:2014}. Then $\Delta t 
=M_T/\dot{M}=2.05$ sec. With the rates reported in table \ref{energytable} this implies a 
total number of counts of $\sim 30750$ for a BH spin $a=0$ and $\sim 48300$ when $a=0.8$. On 
the other hand, using an efficiency of $\varepsilon=0.057$ when $a=0$ and $\varepsilon=0.12$ 
for $a=0.8$ we find $\Delta t \sim 2.3$ sec for both spin values. This results in a total 
number of counts of 34900 and 55200 for $a=0$ and $a=0.8$ respectively.  
%Note, 
%however, that the rates reported in table \ref{energytable} for the C models correspond to a 
%steady state, while the results we report for the J models are based on one snapshot. This 
%means that there will be some extra contribution to the total number of events at later 
%times. However, as the neutrino surfaces shrink and the temperatures and luminosities 
%decrease with time, we expect that contribution to be small.
%================================================================================================

\section{Angular dependence}
\label{angulardependence}

So far we have considered the detection point to be at 10 kpc on the $z$-axis, perpendicular 
to the equatorial plane of the axisymmetric torus. However, given the mass distribution of 
the accreted matter it is natural to expect that neutrino fluxes, rates and any other 
quantity related to them will change according to the observer's inclination. From 
one side, the inclination changes the measured solid angle in the observer's sky. 
Furthermore, neutrinos bend their trajectories due to strong gravity, and regions of the 
neutrino surface that are invisible in Newtonian gravity could be detected in curved space. 
On the other hand, some neutrinos that could reach the observer, thanks to the gravitational 
lensing, can be re-absorbed if on their path to the observer they
enter an opaque region. To 
address these points we locate the observer at $r_o=5000$ km away from the BH. This is a 
reasonable distance that once re-scaled to 10 kpc roughly reproduces the rates found above; 
it is far enough for our treatment of the geodesics in the Schwarzschild metric still to be 
valid, but close enough to allow us study the effect that the inclination has on the solid 
angle apparent to the observer. A direct calculation at 10 kpc will not shed light 
on the intricacies of the apparent solid angle $\Omega_o$.

We estimate the number of neutrinos observed per unit energy, per unit area, per second as
\begin{equation}
\frac{dN}{dA^odt^{o}}=\int dE^{o} \phi^{eff}(E^{o}),
\end{equation}
with the effective flux as in eq. \ref{effflux}.
The observed number of neutrinos per second, luminosities, and average energies are as in 
Eq. 5-7 but evaluated in the observer's frame (the superscript changed by ``$o$``). In this 
case the integral over the area is $4\pi r^2_{o}$. The integral over the solid angle 
$d\Omega_o=\sin\xi d\xi d\alpha$ corresponds to the solid angle subtended by the neutrino 
surface and seen at $r_o$. The observer is now closer to the BH and we need to find $\xi$ by 
explicitly solving the neutrino null geodesics equation. In the Schwarzschild metric this is
\begin{equation}
\left[\frac{1}{r^2}\left(\frac{dr}{d\varphi}\right)\right]^2+\frac{1}{r^2}\left(1-r_s/r\right)=\frac{1}{b^2},
\end{equation}
where $b$ is, as before, the neutrino's impact parameter, and origin of the spherical coordinates 
is centered in the BH. The integration of this equation goes from the emission points on 
the neutrino surface $(r_{e},\varphi_{e})$ to the observation point $(r_{o},\varphi_{o})$. 
Once $b$ is found then $\xi$ (and the solid angle) is given by
\begin{equation}
b=\frac{r_{o} \sin\xi}{\sqrt{1-r_s/r_{o}}}.
\label{xi}
\end{equation}
Neutrinos emitted from a particular point could travel around the BH several times: a 
neutrino emitted close to the BH can fall into it, reach the exterior with some deflection 
(first order image), or rotate around the BH spending more time near the horizon and finally 
escape to reach the observer (higher order image). In this study we limit ourselves to first 
order images. Higher orders will also contribute, though the contribution is expected to be 
subdominant for distant observers and for the BH mass considered here (not the case, 
however, for massive BHs and closer observers \cite{Bozza:2007gt}).

\begin{figure}[ht]
\begin{center}
\includegraphics[trim= 0.in 0.1in 0in 0.1in, clip,width=3.3 in]{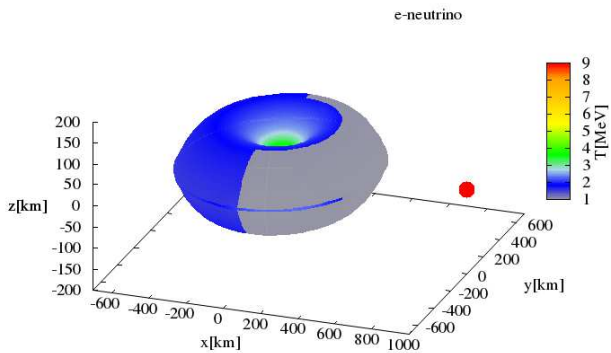}
\caption{Electron neutrino surfaces considered in the fluxes calculation shown in figure \ref{flux}. The full torus assumes
matter around the disk is neutrino transparent. The upper half of the torus will
be detected by an observer in the $z$-axis. The grey right half would be seen by an observer 
in the $z=0$ plane (red dot and not to scale) if the torus is opaque. The curved line on the torus shows the intersection of the its surface
with the $z=0$ plane. The torus model is
the steady-state Ca with a spin $a$=0.95 and accretion rate 5$M_\odot$/s.}
\label{thickscketch}
\end{center}
\end{figure}

As mentioned above, after neutrinos have left the neutrino surface they can pass again 
through an optically thick region on their way to the observer. The apparent solid angle is 
then not only affected by gravitational lensing but also by neutrino interactions with 
matter. To study the effect on neutrino fluxes due to changes in the apparent solid angle 
only we assume the torus is optically thin. By this we mean: once the neutrinos have 
decoupled from the optically thick region (at the neutrino surface), they do not 
re-encounter thick regions. In such case all the neutrino surface shown in figure 
\ref{thickscketch} emits. This provides us with upper limits of our estimates. We contrast 
these results with two totally thick disks: one as seen by an observer on the $z$-axis 
(upper half torus, $z>0$); and one where the observer is located on the $z=0$ plane, noted 
by the red dot (not to scale), who will detect the grey surface (roughly half right torus) 
which starts at largest distance in the $z$-axis of the neutrino surface, see figure 
\ref{thickscketch}. For both thick tori we are assuming that the parts of neutrino surface 
opposite to the obsever are undetected. This will lead to underestimates as neutrinos from 
those parts can actually reach the observer.

\begin{figure}
\begin{center}
\includegraphics[width=0.45\textwidth]{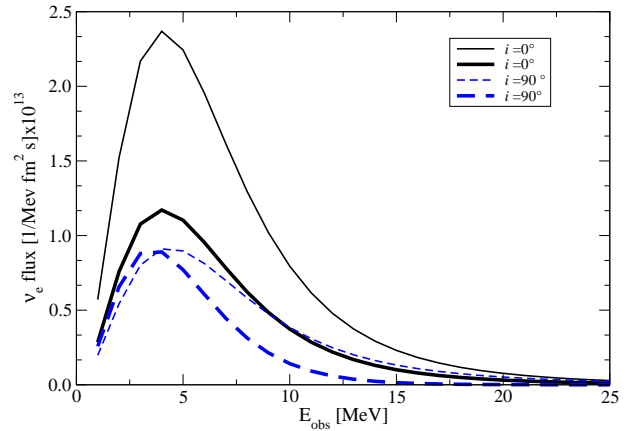}
\caption{Electron neutrino fluxes as seen by an observer at 5000 km from a BH of $3M_\odot$,
 and polar inclinations of $i=0^\circ$ and 90$^\circ$. The torus model is
the steady-state Ca with a spin $a$=0.95 and accretion rate 5$M_\odot$/s. Thin lines correspond to neutrino transparent 
tori while thick lines to opaque ones.}
\label{flux}
\end{center}
\end{figure}

Figure \ref{flux} shows the electron neutrino fluxes at 5000 km from the center of the 
torus. The solid lines correspond to an observer's inclination $i=0^\circ$, with respect to 
$z$-axis of the torus, and the dashed lines to $i=90^\circ$. The thick lines assume the 
torus is optically thick while the thin lines correspond to an optically thin torus. In this 
way, an observer located at $i^\circ=0$ will register fluxes that are emitted from the upper 
half of the neutrino surface only. If the torus is optically thin neutrinos will reach the 
observer from the entire neutrino surface (including the lower half). Then at $i=0^\circ$ 
the thin fluxes roughly double the thick ones (compare black solid lines). At an inclination 
of 90$^\circ$ (blue dashed lines), which is on the equatorial plane, the situation is 
different. If the torus is totally thick an observer will register fluxes emitted from the 
grey right half torus in figure \ref{thickscketch} only. Then the hottest inner parts of the 
disk won't contribute. In the thin case neutrinos are emitted everywhere from the neutrino 
surface, enhancing the fluxes. Furthermore neutrinos emitted from the opposite half of the 
neutrino surface can also reach the observer due to strong deflection with lower chances of 
being reabsorbed (their geodesics do not cross the highest density regions). Additionally at 
this angle, the Doppler effect overcompensates the gravitational redshift, due to the 
angular velocity carried by the disk. The final result are the larger fluxes at higher 
energies described by the thin blue dashed line.
 
Based on fluxes obtained at different inclinations, with respect to the $z$-axis, we 
estimate electron antineutrino average energies, luminosities and detection rates at SuperK. 
The results are shown in Figure 5, for a torus of accretion rate 5$M_\odot$/s and for BH 
spins $a=0$ and $a=0.95$ (C0 and Ca models). The quantities are calculated at 5000 km 
from the BH and the detection rates have been rescaled to 10kpc. Red lines correspond to a 
BH with spin $a=0.95$ (indicated with $a$ in the legend), while estimates corresponding to a 
BH with zero spin are represented by black lines. If the torus is neutrino transparent (i.e. 
assuming they don't re-encounter an optically thick region after they leave the neutrino 
surface) the full neutrino surface is considered and the results are represented by solid 
thin lines. We contrast these results by considering emission of the upper half only, i.e. 
above the equatorial plane, of the neutrino surface (dotted-dashed thick lines).

\begin{figure}[ht]
\begin{center}
\includegraphics[width=0.48\textwidth]{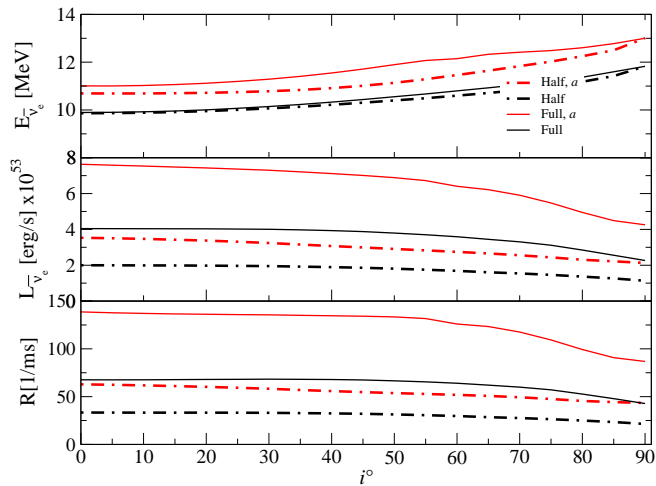}
\caption{Electron antineutrino energies, luminosities and detection rates $R$ at SuperK as function of the inclination
of the observer with respect to the $z$-axis. The observer is at 5000 km from a BH of $3M_\odot$ and the torus' accretion
rate is 5$M_\odot$/s. The rates $R$ are rescaled at 10 kpc. 
Red and black lines correspond to spins $a=0.95$ and $0$ respectively. Solid lines assume the tori are neutrino transparent 
while in the calculation showed by the dotted-dashed lines we have used the upper-half of the disk only.}
\label{evsi}
\end{center}
\end{figure}

As the inclination increases the solid angle subtended by the neutrino surface decreases, 
producing a reduction in the luminosities and rates. The average energy, estimated as the 
ratio between the luminosity and the number of neutrinos per second, becomes insensitive to 
this reduction. Its behavior is mainly affected by the energy shifts. At zero inclination 
the neutrino energies observed from different parts of the neutrino surface are all 
similarly redshifted: gravitational redshift dominates over Doppler shifts. In this case 
the linear velocity of an element of emitting matter  is 
perpendicular to the observer's. As the inclination increases the linear velocity of the 
disk becomes parallel to the observer's. Then energies of half the torus are redshifted 
and the other half appear blueshifted. Thus the shifts in energies are totally different 
from the shifts in the perpendicular direction. The overall effect is a maximum increase in 
the observed average energies of about 2 MeV for $a=0.95$ and 1 MeV for $a=0$ from $i=0$ to 
$i=90$ deg. Considering a transparent (full) torus has the obvious consequence of having 
roughly twice the luminosities and rates than the corresponding quantities for a half torus. 
The small differences in the average energies are due to the fact that the lower parts (for 
$z<0$) of the full neutrino surface are farther from the observer who registers different 
redshifts for neutrinos coming from that area compared to the ones coming from the upper 
half. When the inclination is 90 degrees the number luminosities and luminosities of the 
full torus are twice half the disk and the average energies are the same. 
%A discussion of 
%the increment of the lumniosites, rates and energies with the BH spin has been made above. 
%We complement this by pointing out that 
The differences in average energies when a full or 
half torus are considered are larger for spinning BHs than for the $a=0$ case. This is 
because the energy shift depends on the BH spin and therefore the higher the spin the larger 
the differences in the average energies between the full and half tori.

%====================================================================================================
\section{Summary and Conclusions}
\label{conclusions}

Collapsars, BH-NS and NS-NS mergers can evolve into a BH surrounded by an accretion disk. 
The physics of these objects is crucial in our understanding of gravitiational waves, 
gamma ray bursts and the 
production of heavy neutron-rich nuclei. The structure of the BH affects the thermodynamic 
properties of the disk leading to changes in the neutrino surfaces, their fluxes, and 
consequently in the setting of the neutron to proton fraction, and annihilation rates. We 
have studied the influence that the BH spin and the location of a distant observer have on 
several neutrino properties using a ray-tracing technique in general relativity. For the 
torus model we have used two different approaches: 1) an extension to 3D of a one 
dimensional fully relativistic steady state disk and 2) a time-dependent axisymmetric 
hydrodynamical torus. For both models we found that the spin of the BH strongly affects the 
behavior of the neutrino surfaces. Regardless of the flavor, the neutrino surface of an 
accreting torus around a spinning BH has, when compared to the non-spinning BH, higher 
neutrino temperatures and larger neutrino surfaces. These characteristics lead to higher 
neutrino luminosities and average energies. More luminous neutrinos would annihilate at a 
higher rate and therefore, rotating BHs with total neutrino luminosities $\sim 10^{54}$ are 
a more likely short gamma ray burst source than steady ones (in agreement with 
\cite{SetiawanBHmerger}).

Our estimates for neutrino counts per second at SuperK, assuming the source is at 10 kpc, 
are larger for disks around spinning BHs, and these detection rates are at least a factor of 
10 larger than supernova detection rates, regardless of the torus model. This due to the 
larger temperatures. The total number of counts and the duration of the signal depend on the 
efficiency of converting gravitational energy into neutrino energy. Spinning BHs are more 
efficient and this is reflected in the larger counts we predict compared to the tori around 
static BHs. For constant accretion rates we estimate signals lasting tenths of seconds, 
whereas the number of events varies from few to several thousands (increasing with the 
accretion rate).

The fluxes, luminosities, average energies and detection rates depend on the angle that the 
disk describes in the observer's sky. At larger inclinations the fluxes, luminosities, and 
detection rates decrease. This is due to the fact that the apparent area of the neutrino 
surface becomes smaller with increasing angles. Due to Doppler effects, included already in 
the energy shifts, the average energies increase with the inclination of the observer.

As the possibly observable quantities depend on the apparent emission area to the observer, 
it is relevant to ask: What is the appropriate neutrino surface for a given observer's 
inclination? Here we assumed that once the neutrinos decouple from matter, at the last 
points of scattering, they do not pass through an optically thick region again. This however 
could lead to overestimates and so we contrast these results with a case where only the upper 
half of the neutrino surface (on the $z>0$) emits. For example, in the case of an observer 
located on the axis perpendicular to the plane of the disk (here $z$-axis) it is reasonable 
to assume that neutrinos located in the lower half will likely be reabsorbed as they are 
back to the high density regions \lq\lq inside\rq\rq\ the neutrino surface. Then at zero inclination 
our estimates from the upper half of the torus seem reasonable, whereas the full disk 
provides us with upper limits. However, geodesics coming from the lower part could indeed 
reach the observer due to the deflection of neutrino trajectories. This effect is stronger 
as the observer's inclination is increased. For an observer located on the equatorial plane 
($i=90^\circ$) neutrinos emitted from the farthest (opposite) parts of the torus leave the 
disk with trajectories tangential to the disk and have fewer chances to be re-absorbed. Then 
gravitational effects would allow us to \lq\lq see\rq\rq\ the opposite side of the neutrino surface. 
Similarly, the hottest neutrinos emitted near the BH can be strongly deflected and reach the 
observer contributing to the total flux. At this inclination, considering only a part of the 
neutrino surface as the emitting area would lead to underestimates. We found that the 
average energies are not largely affected by these considerations compared to the effects on 
luminosities and detection rates. The latter could be overestimated by roughly a factor 
of two if the tori are assumed transparent for zero inclination, and perhaps by a larger 
factor for observers near the equatorial plane of the disks. We will leave for future work a 
detailed study of the effects of reabsorption of the neutrino geodesics.

The results presented here also motivate future studies on the effects of the BH spin on the 
synthesis of elements. Neutrinos set the electron fraction and any changes in the neutrino 
fluxes will be reflected on the element abundances, particularly in the wind nucleosynthesis. 
Furthermore, the properties of the torus 
outflows will change at different inclinations. The gravitational effects will be much more 
prominent as the observer, in this case the outflowing matter, is closer to the BH than the 
observers considered here. We will also address this in future work.

\acknowledgements
This work was partially supported by the Natural Sciences and
Engineering Research Council of Canada (NSERC)(OLC) and by U.S. DOE Grants
No. DE-FG02-02ER41216(GCM), DE-SC0004786(GCM) and DE-SC0013039(RS).

%%%%%%%%%%%%%%%%%%%%%%%%%%%%%%%%%%%%%%%%%%%%%%%%%%%%%%%%%%%%%%%%%
\vfill\eject


\begin{thebibliography}{99}

\bibitem {MacFadyen1999} A. MacFadyen and S. E Wossley, ApJ {\bf 524} (1999) 262.
\bibitem{Taniguchi2005} K. Taniguchi, T. W. Baungarte, J. A. Faber and S. L. Shapiro, PRD {\bf 72} (2005), 044008. 
\bibitem{Lee1999} W. Kluzniak and W. H. Lee,  MNRAS {\bf 308} (1999) 780.
\bibitem{Rosswog} S. Rosswog, arXiv:astrop-ph/0508138.
\bibitem{Lattimer1974} J. M. Lattimer and D. N. Schramm, ApJ,{\bf 192}(1974) L145.
\bibitem{Lattimer1976} J. M. Lattimer and D. N. Schramm, ApJ, {\bf 210}(1976) 549.

%\cite{Surman:2008qf}
\bibitem{Surman:2008qf} 
  R.~Surman, G.~C.~McLaughlin, M.~Ruffert, H.-T.~Janka and W.~R.~Hix,
  %``r-Process Nucleosynthesis in Hot Accretion Disk Flows from Black Hole - Neutron Star Mergers,''
  Astrophys.\ J.\  {\bf 679}, L117 (2008)

%\cite{Fernandez:2014cna}
\bibitem{Fernandez:2014cna} 
  R.~Fern\'{a}ndez, D.~Kasen, B.~D.~Metzger and E.~Quataert,
  %``Outflows from accretion discs formed in neutron star mergers: effect of black hole spin,''
  Mon.\ Not.\ Roy.\ Astron.\ Soc.\  {\bf 446}, 750 (2015)

\bibitem{SetiawanBHmerger} S. Setiawan, M. Ruffert and H.-Th. Janka, Astron. Astrophys., {\bf 458} (2006), 553.


\bibitem{RuffertGRB-BH} M. Ruffert, H.-Th. Janka, Astron. Astrophys.,  {\bf 344} (1999) 573.
\bibitem{Popham1999} Robert Popham, S. E. Woosley and Chris Fryer, ApJ {\bf 518} (1999) 356.

  \bibitem{Nakamura:2013bza}
  K.~Nakamura, S.~Harikae, T.~Kajino and G.~J.~Mathews,
  %``r-process nucleosynthesis in the neutrino-heated relativistic collapsar jet model for gamma-ray bursts,''
  PoS NICXII {\bf } (2012) 216.
\bibitem{Nagatakicounts} Shigehiro Nagataki and Kazunori Kohri, Prog. Theor. Phys., {\bf 108} (2002)789.

%\cite{McLaughlin:2006yy}
\bibitem{McLaughlin:2006yy} 
  G.~C.~McLaughlin and R.~Surman,
  %``Supernova Neutrinos: The Accretion Disk Scenario,''
  Phys.\ Rev.\ D {\bf 75}, 023005 (2007)


%\cite{Caballero:2009ww}
\bibitem{Caballero:2009ww} 
  O.~L.~Caballero, G.~C.~McLaughlin, R.~Surman and R.~Surman,
  %``Detecting neutrinos from black hole neutron stars mergers,''
  Phys.\ Rev.\ D {\bf 80}, 123004 (2009)
 
    \bibitem{Foucart:2015vpa} 
  F.~Foucart {\it et al.},
  %``Post-merger evolution of a neutron star-black hole binary with neutrino transport,''
  Phys.\ Rev.\ D {\bf 91}, no. 12, 124021 (2015)
  
  \bibitem{Foucart:2014nda} 
  F.~Foucart {\it et al.},
  %``Neutron star-black hole mergers with a nuclear equation of state and neutrino cooling: Dependence in the binary parameters,''
  Phys.\ Rev.\ D {\bf 90}, 024026 (2014)
  
  \bibitem{Deaton:2013sla} 
  M.~B.~Deaton {\it et al.},
  %``Black Hole-Neutron Star Mergers with a Hot Nuclear Equation of State: Outflow and Neutrino-Cooled Disk for a Low-Mass, High-Spin Case,''
  Astrophys.\ J.\  {\bf 776}, 47 (2013)
  
    \bibitem{Sekiguchi:2015dma} 
  Y.~Sekiguchi, K.~Kiuchi, K.~Kyutoku and M.~Shibata,
  %``Dynamical mass ejection from binary neutron star mergers: Radiation-hydrodynamics study in general relativity,''
  Phys.\ Rev.\ D {\bf 91}, no. 6, 064059 (2015).
\bibitem{Palenzuela:2015dqa} 
  C.~Palenzuela, S.~L.~Liebling, D.~Neilsen, L.~Lehner, O.~L.~Caballero, E.~OâConnor and M.~Anderson,
  %``Effects of the microphysical Equation of State in the mergers of magnetized Neutron Stars With Neutrino Cooling,''
  Phys.\ Rev.\ D {\bf 92}, no. 4, 044045 (2015). 
  
    \bibitem{Etienne:2008re} 
  Z.~B.~Etienne, Y.~T.~Liu, S.~L.~Shapiro and T.~W.~Baumgarte,
  %``General relativistic simulations of black-hole-neutron-star mergers: Effects of black-hole spin,''
  Phys.\ Rev.\ D {\bf 79}, 044024 (2009)
  
  \bibitem{Foucart:2010eq} 
  F.~Foucart, M.~D.~Duez, L.~E.~Kidder and S.~A.~Teukolsky,
  %``Black hole-neutron star mergers: effects of the orientation of the black hole spin,''
  Phys.\ Rev.\ D {\bf 83}, 024005 (2011)
  
\bibitem{Beloborodovcross}Wen-Xin Chen and Andrei M. Beloborodov, ApJ {\bf657} (2007) 383.
\bibitem{Zalamea:2008dq} 
  I.~Zalamea and A.~M.~Beloborodov,
  %``Efficiency of Neutrino Annihilation around Spinning Black Holes,''
  AIP Conf.\ Proc.\  {\bf 1133}, 121 (2009)
  \bibitem{Birk07:annhihilation}R. Birkl, M. A. Aloy, H.-Th. Janka and E. Muller, Astron. Astrophys., {\bf 463} (2007) 51.
    
   \bibitem{Liu:2015lfa} 
  T.~Liu, S.~J.~Hou, L.~Xue and W.~M.~Gu,
  %``Jet Luminosity of Gamma-ray Bursts: The Blandford-Znajek Mechanism versus the Neutrino Annihilation Process,''
  Astrophys.\ J.\ Suppl.\  {\bf 218}, no. 1, 12 (2015)
  
  
  \bibitem{Harikae:2009yt} 
  S.~Harikae, K.~Kotake and T.~Takiwaki,
  %``Neutrino Pair Annihilation in Collapsars: Ray-Tracing Method in Special Relativity,''
  Astrophys.\ J.\  {\bf 713}, 304 (2010)  
  

%\cite{Surman:2013sya}
\bibitem{Surman:2013sya} 
  R.~Surman, O.~L.~Caballero, G.~C.~McLaughlin, O.~Just and H.~T.~Janka,
  %``Production of $^{56}Ni$ in black hole-neutron star merger accretion disc outflows,''
  J.\ Phys.\ G {\bf 41}, 044006 (2014)


%\cite{Caballero:2011dw}
\bibitem{Caballero:2011dw} 
  O.~L.~Caballero, G.~C.~McLaughlin and R.~Surman,
  %``Neutrino Spectra from Accretion Disks: Neutrino General Relativistic Effects and the Consequences for Nucleosynthesis,''
  Astrophys.\ J.\  {\bf 745}, 170 (2012)

%\cite{Malkus:2012ts}
\bibitem{Malkus:2012ts} 
  A.~Malkus, J.~P.~Kneller, G.~C.~McLaughlin and R.~Surman,
  %``Neutrino oscillations above black hole accretion disks: disks with electron-flavor emission,''
  Phys.\ Rev.\ D {\bf 86}, 085015 (2012)
  [arXiv:1207.6648 [hep-ph]].

%\cite{Malkus:2014iqa}
\bibitem{Malkus:2014iqa} 
  A.~Malkus, A.~Friedland and G.~C.~McLaughlin,
  %``Matter-Neutrino Resonance Above Merging Compact Objects,''
  arXiv:1403.5797 [hep-ph].
 

\bibitem{SK} S. Fukuda et al., Nucl. Inst. Meth. Phys. A, {\bf 501} (2003) 418.


\bibitem{Just:2014} O. Just et al. submitted to MNRAS (2014) arXiv:1406.2687
\bibitem{BHpote}B. Paczynski and P. Wiita, Astron. Astrophys., {\bf 88} (1980) 32.
\bibitem{Artemova}I. V. Artemova, G. Bjornsson and I. D. Novikov, ApJ {\bf 461} (1996) 565.

\bibitem{Caballerogravity} O. L. Caballero, G. C. MacLaughlin and R. Surman, http://arxiv.org/abs/1410.7663.
\bibitem{Perego:2014fma} 
  A.~Perego, S.~Rosswog, R.~M.~Cabezón, O.~Korobkin, R.~Käppeli, A.~Arcones and M.~Liebendörfer,
  %``Neutrino-driven winds from neutron star merger remnants,''
  Mon.\ Not.\ Roy.\ Astron.\ Soc.\  {\bf 443}, no. 4, 3134 (2014)
  [arXiv:1405.6730 [astro-ph.HE]].

\bibitem{Shapiro} S. L. Shapiro and S. A Teukolsky, {\it Black Holes, White Dwarfs and Neutron Stars}, Wiley, New York, 1983.
  \bibitem{Bozza:2007gt} 
  V.~Bozza and G.~Scarpetta,
  %``Strong deflection limit of black hole gravitational lensing with arbitrary source distances,''
  Phys.\ Rev.\ D {\bf 76}, 083008 (2007).
  
    
\end{thebibliography}
\end{document}